\documentclass[3p,authoryear]{elsarticle}
\journal{Planetary \& Space Science}

\usepackage{amssymb}

\usepackage{lineno}
\usepackage{textcomp}
\usepackage{doi}
\usepackage{color}
\usepackage[usenames,dvipsnames]{xcolor}
\usepackage{ulem}

\usepackage{hyperref}
\hypersetup{colorlinks=true, urlcolor=blue}

\begin{document}

\begin{frontmatter}



\title{Variegation and space weathering on asteroid 21 Lutetia\tnoteref{label1}\tnoteref{label2}}
\tnotetext[label1]{\doi{10.1016/j.pss.2015.06.018}}
\tnotetext[label2]{\copyright 2017. This manuscript version is made available under the CC-BY-NC-ND 4.0 licence\\ \url{https://creativecommons.org/licenses/by-nc-nd/4.0/}}


\author[DLR]{S.E.~Schr\"oder\corref{cor1}}
\ead{stefanus.schroeder@dlr.de}
\author[TUB]{H.U.~Keller}
\author[DLR]{S. Mottola}
\author[DLR]{F. Scholten}
\author[DLR]{F. Preusker}
\author[DLR]{K.-D.~Matz}
\author[DLR]{S. Hviid}

\cortext[cor1]{Corresponding author}

\address[DLR]{Deutsches Zentrum f\"ur Luft- und Raumfahrt (DLR), 12489 Berlin, Germany}
\address[TUB]{Institut f\"ur Geophysik und extraterrestrische Physik (IGEP), Technische Universit\"at Braunschweig, 38106 Braunschweig, Germany}

\begin{abstract}

During the flyby in 2010, the OSIRIS camera on-board Rosetta acquired hundreds of high-resolution images of asteroid Lutetia's surface through a range of narrow-band filters. While Lutetia appears very bland in the visible wavelength range, \citet{Mg12} tentatively identified UV color variations in the Baetica cluster, a group of relatively young craters close to the north pole. As Lutetia remains a poorly understood asteroid, such color variations may provide clues to the nature of its surface. We take the color analysis one step further. First we orthorectify the images using a shape model and improved camera pointing, then apply a variety of techniques (photometric correction, principal component analysis) to the resulting color cubes. We characterize variegation in the Baetica crater cluster at high spatial resolution, identifying crater rays and small, fresh impact craters. We argue that at least some of the color variation is due to space weathering, which makes Lutetia's regolith redder and brighter.

\end{abstract}

\begin{keyword}
Asteroid Lutetia \sep Spectrophotometry \sep Opposition effect \sep Space weathering
\end{keyword}

\end{frontmatter}

%\begin{linenumbers}

\section{Introduction}
\label{sec:introduction}

After 2867 {\v S}teins, 21 Lutetia was the second asteroid visited by the Rosetta spacecraft. In July 2010, the on-board OSIRIS camera \citep{K07} acquired hundreds of high-resolution images of the northern hemisphere through a series of narrow-band filters that span a wavelength range from the UV to the near-IR. The first OSIRIS results from the encounter were described by \citet{S11}. The shape of Lutetia is roughly an ellipsoid with dimensions of $121 \times 101 \times 75$~km, and a detailed shape model based on the images validates an earlier model based on ground-based observations \citep{C12}. OSIRIS determined a visual geometric albedo of $0.19 \pm 0.01$, confirming the estimate of $0.20 \pm 0.03$ by \citet{S08}. Lutetia is considered a somewhat enigmatic body. Due to its relatively high albedo it was classified with the M-type asteroids, a group thought to harbor metallic asteroids \citep{L10,W10}. However, its reflectance spectrum in the near- and mid-infrared resembles more that of CV and CO carbonaceous chondrites \citep{B08,L09,B10}. Adding to the puzzle is Lutetia's radar albedo, which is not consistent with that of typical metallic asteroids like Kleopatra or Psyche, but also considerably higher than the average for C-type asteroids \citep{S08}. Despite the wealth of data, the flyby did not conclusively establish Lutetia's composition. Based on data from the on-board VIRTIS spectrometer, \citet{C11} speculated that Lutetia is metal-rich and composed of mafic silicate minerals with a low iron abundance. The authors cited its flat spectrum as evidence for absence of spectral changes due to space weathering. \citet{B12} suggested that the asteroid surface is a mixture of materials considered incompatible, specifically carbonaceous and enstatite chondrites, perhaps as a consequence of large impacts. \citet{St11} established that Lutetia's reflectance spectrum from 120 to 800~nm is best matched by that of an EH5 chondrite, an iron-free member of the orthopyroxene mineral family. Spectral features observed by the Alice UV spectrometer hint at the presence of H$_2$O and SO$_2$ frost. \citet{W12} suggested that Lutetia is partially differentiated, with a primitive chondritic crust overlaying a metallic core.

\begin{figure}
\centering
\includegraphics[width=\textwidth]{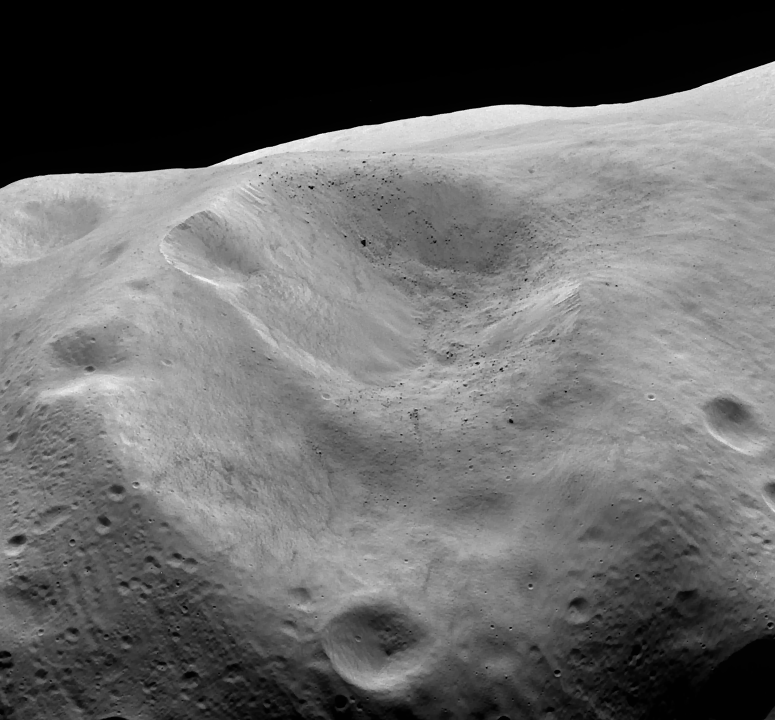}
\caption{The crater complex in the Baetica region in which color variations were tentatively identified by \citet{Mg12} ({\it Orange} image 15:44:41, sharpened as described in \citealt{S10}).}
\label{fig:Baetica_Orange}
\end{figure}

Impacts on a differentiated asteroid can give rise to considerable surface variegation, as the example of Vesta shows \citep{R12}. Indeed, ground-based observations prior to the Rosetta flyby found evidence for surface heterogeneity \citep{N07,L10,P10}. \citet{Mg12} carefully calibrated the OSIRIS color images and made a first analysis of the surface variegation using a rubber-sheet algorithm to orthorectify the images. They found evidence for weak surface variegation in the UV that is restricted to landslides in a crater complex in the Baetica region (Fig.~\ref{fig:Baetica_Orange}), and speculated on a connection with differences in freshness and/or grain size of the soil. The North Polar Crater Cluster in the center of Baetica was formed by at least three major impacts and a number of smaller ones \citep{M12,T12}. The presence of fresh landslides and boulders in and around the complex attests to its youth \citep{K12}. We take the analysis of OSIRIS color images one step further and orthorectify the images by projecting them onto a shape model \citep{P12} while improving the camera pointing reconstruction. This allows us to study spectral variability at a much higher spatial resolution than before. Furthermore, we attempt to compensate for the effects of illumination and local topography by photometrically correcting the images. Alternatively, we apply a principal component analysis (PCA) to isolate subtle color variations by eliminating apparent brightness variations due to topography and albedo. Our improved treatment provides additional insight into the nature of Lutetia's surface.

\section{Data}
\label{sec:data}

The OSIRIS Narrow Angle Camera (NAC) images used in this paper are calibrated to reflectance (radiance factor, $I/F$) and were retrieved from the Planetary Science Archive \citep{H11}. We concentrate our analysis on two sets of images that we refer to as ``set~1'' and ``set~2'' (Table~\ref{tab:data_sets}). In both sets the images were taken through a sequence of color filters in rapid succession, such that the changes in phase angle and distance to the asteroid are small within a set. Set~1 was acquired at a lower phase angle and larger distance than set~2. The phase angle ranges of the images in set~1 and 2 are $2.7^\circ$-$4.2^\circ$ and $15.9^\circ$-$21.4^\circ$, respectively. The spatial resolution of the images in set~2 is about twice that for set~1. All available NAC color filters were used; their characteristics\footnote{Note that the filter names are merely descriptive and not necessarily correct. For example, the {\it Orange} filter is centered on a wavelength that our eyes would perceive as red, and the {\it Red} filter observes in the near-infrared, invisible to our eyes.} are listed in Table~\ref{tab:filters}. OSIRIS images do not have a unique identifying number (although the full file name is unique), so we classify them according to imaging time reported in the file name. All images we analyze in this paper were acquired on July 10, 2010, and we refer to individual images by their file name time in the format ``hh:mm:ss''. As all images were taken at different distances to Lutetia and viewing geometries, we orthorectified them to allow direct comparison. This was done by mapping to a Lambert azimuthal projection with $90^\circ$ latitude (north pole) at the center, using the shape model and improved orientation data (orbit position and pointing) that were derived by stereo-photogrammetric analysis \citep{P12}.

\section{Color variations}

\subsection{Color composites}
\label{sec:color_composites}

\begin{figure}
\centering
\includegraphics[width=7cm]{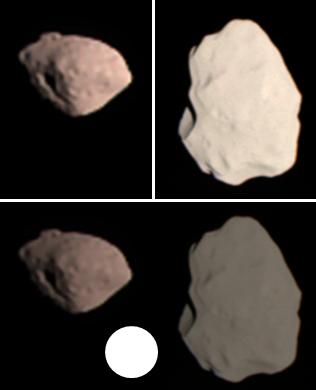}
\caption{Asteroids {\v S}teins (left) and Lutetia (right) as observed by OSIRIS NAC in true color, assuming solar illumination and CIE color matching functions from \citet{V78}. The top images show both at maximum brightness, that is, as how they would appear to our eyes after letting them get accustomed to the brightness of the asteroids. The bottom images show both at their correct relative brightness at zero phase angle, where a perfectly reflecting Lambertian disk would be perceived as white. Note that the actual asteroid images shown here were not acquired at zero phase angle. True color was calculated using a {\v S}teins spectrum from \citet{F08} and a Lutetia spectrum from \citet{P10}.}
\label{fig:true_color}
\end{figure}

\begin{figure}
\centering
\includegraphics[width=\textwidth]{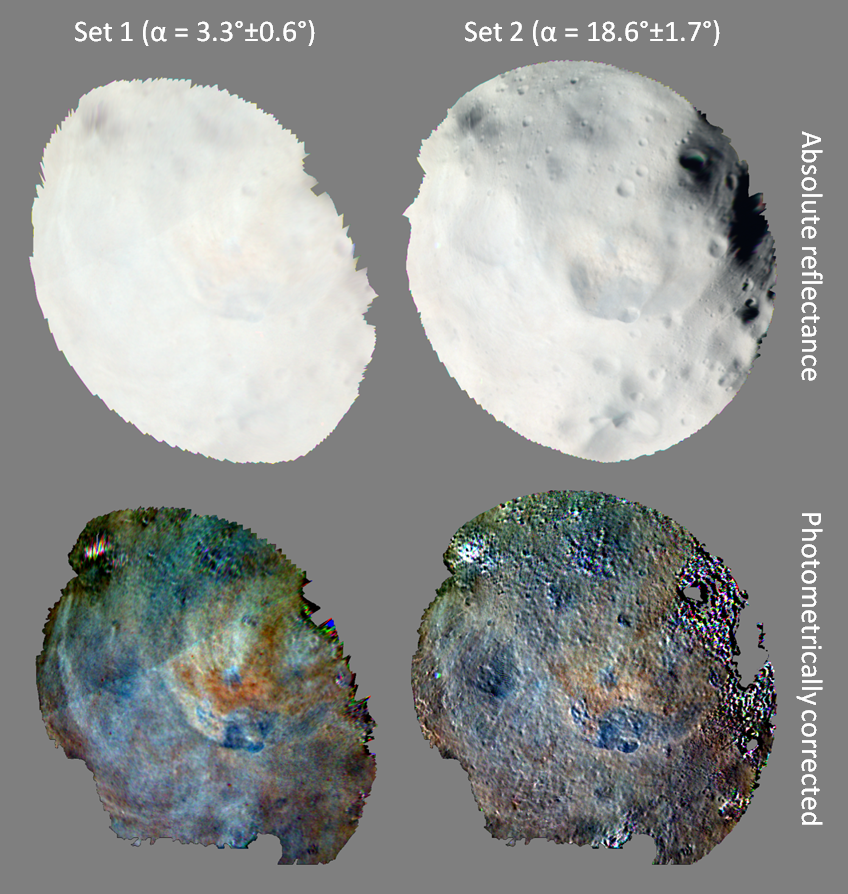}
\caption{Color composites of the orthorectified images of sets~1 (left column) and 2 (right column). The top row are color composites with the absolute reflectance in the {\it Orange}, {\it Near-UV}, and {\it Far-UV} filters in the R, G, and B channels, respectively. Reflectance zero is black and the maximum reflectance is white. The bottom row are contrast-enhanced color composites of the same images photometrically corrected with the \citet{H02} model.}
\label{fig:color}
\end{figure}

In the visible wavelength range Lutetia appears dark and gray. Figure~\ref{fig:true_color} shows Lutetia in true colors, that is, as it would appear to our eyes, both by itself as well as next to {\v S}teins. The surface appears devoid of color variations. However, minor variegation was tentatively detected by \citet{Mg12}. To study these small variations we enhance the images by means of photometric correction, i.e.\ by removing the brightness variations due to shading by topography. Figure~\ref{fig:color} shows color composites before and after photometric correction using the best-fit \citet{H02} model with $w = 0.218$, $B_{{\rm S}0} = 1.91$, $h_{\rm S} = 0.0502$, $\bar{\theta} = 23.6^\circ$, and Henyey-Greenstein parameter $g = -0.283$ (this model was used for all bands). After investigating different combinations, we determined that the three-filter combination {\it Far-UV}, {\it Near-UV}, and {\it Orange} (see Table~\ref{tab:filters}) exhibits the largest amount of color variation, consistent with \citet{Mg12}. When we simply construct an RGB color composite by assigning the absolute reflectance in the three filters to each color channel, the surface looks rather bland (Fig.~\ref{fig:color}, top row). Subtle color variations are associated with the Baetica region, where the interior of the central crater has an orange tint and a major landslide ({\it Danuvius Labes}\footnote{See http://planetarynames.wr.usgs.gov/Page/LUTETIA/target for names of features on Lutetia as defined by the International Astronomical Union.}) appears slightly blue. This is more obvious in the low phase angle color composite (set~1) for which shading is minimal. These color variations are confirmed when we photometrically correct the images and enhance the contrast by linear scaling (Fig.~\ref{fig:color}, bottom row). The set~1 composite reveals a very colorful Baetica region with an orange/brown interior, two dark blue areas, and several white patches. While the corrected set~2 composite essentially confirms the color variations in the set~1 composite, it has a noisier, rougher appearance. This derives from small-scale artifacts in the \citet{P12} shape model that are amplified when photometrically correcting the images. Also note that the top left part of the corrected composite of set~1 is slightly darker than the bottom right part due to a calibration artifact associated with a different gain for two parts of the CCD.

Underlying our interpretation of the color variations is the assumption that they are not due to the exposure of different mineralogical units in the interior. While Lutetia might be differentiated, at least the upper layer of the crust must be thoroughly homogenized. The Baetica crater cluster is too small and the impacts that led to its formation were too weak to have punched through this layer. That said, the impacts themselves may have introduced different mineralogical units to the surface, as is believed to have happened on asteroid Vesta \citep{McC12,Re12}.

\begin{figure}
\centering
\includegraphics[width=\textwidth]{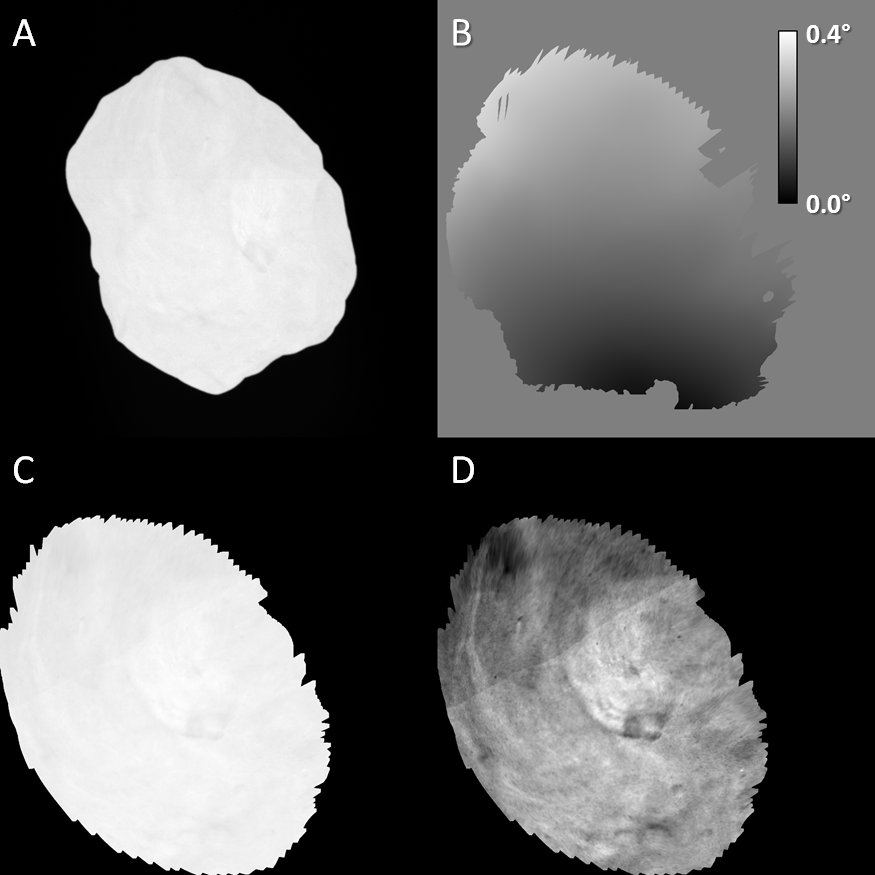}
\caption{Lutetia at opposition in {\it Orange} image 15:26:36. {\bf A}: Calibrated image (level~2). {\bf B}: Map of the phase angle distribution in the orthorectified image. {\bf C}: Orthorectified image displayed with zero reflectance black and maximum reflectance white. {\bf D}: Same, now with the full brightness range (black to white) assigned to the top 15\% of the reflectance scale. B, C, and D have the same projection.}
\label{fig:opposition}
\end{figure}

Interpretation of the variations in the photometrically corrected composites is not straightforward because they also include variations in albedo. To separate these two factors we turn to an image showing predominantly albedo variations for a single color. Image 15:26:36 in Fig.~\ref{fig:opposition} was acquired as part of a sequence of images taken through the {\it Orange} filter to characterize the opposition effect \citep{G56}. At opposition the sun is directly behind the spacecraft, so shadows are hidden and all brightness variations are intrinsic to the surface. If the surface scatters Lommel-Seeliger-like rather than Lambertian at zero phase, such brightness variations are due to variations in albedo rather than topography. This is the case for Lutetia, demonstrated by the simple fact that the asteroid image at opposition is perceived as flat, with no appreciable limb darkening (Fig.~\ref{fig:opposition}A). The phase angle is variable over the disk (by about $0.4^\circ$ for set~1 and $0.9^\circ$ for set~2), and here it ranges from $0.018^\circ$ to $0.346^\circ$ (Fig.~\ref{fig:opposition}B). So we are very close to true opposition over the entire image. Displayed with the full dynamic range, Lutetia shows little contrast at opposition (Fig.~\ref{fig:opposition}C). After enhancing the contrast (Fig.~\ref{fig:opposition}D) we see that the central crater complex in the Baetica region is brighter than Lutetia average, with the exception of the landslide at the bottom of the complex, which is darker than average. The excursions in brightness are about 3\%. Brightness variations over the disk are expected due to the phase angle gradient, but limited to at most a few percent because the opposition effect is blunted by the apparent size of the Sun of $0.2^\circ$. (Note that the aforementioned 1\% gain difference between the two parts of the CCD causes a noticeable change in brightness in Fig.~\ref{fig:opposition}D.) The brightness changes in the Baetica region are on small scales, unrelated to the gradient, and must be due to albedo changes. The {\it Orange} filter has a central wavelength in the red range (Table~\ref{tab:filters}), hence Fig.~\ref{fig:opposition} essentially shows the red albedo.

\begin{figure}
\centering
\includegraphics[width=\textwidth]{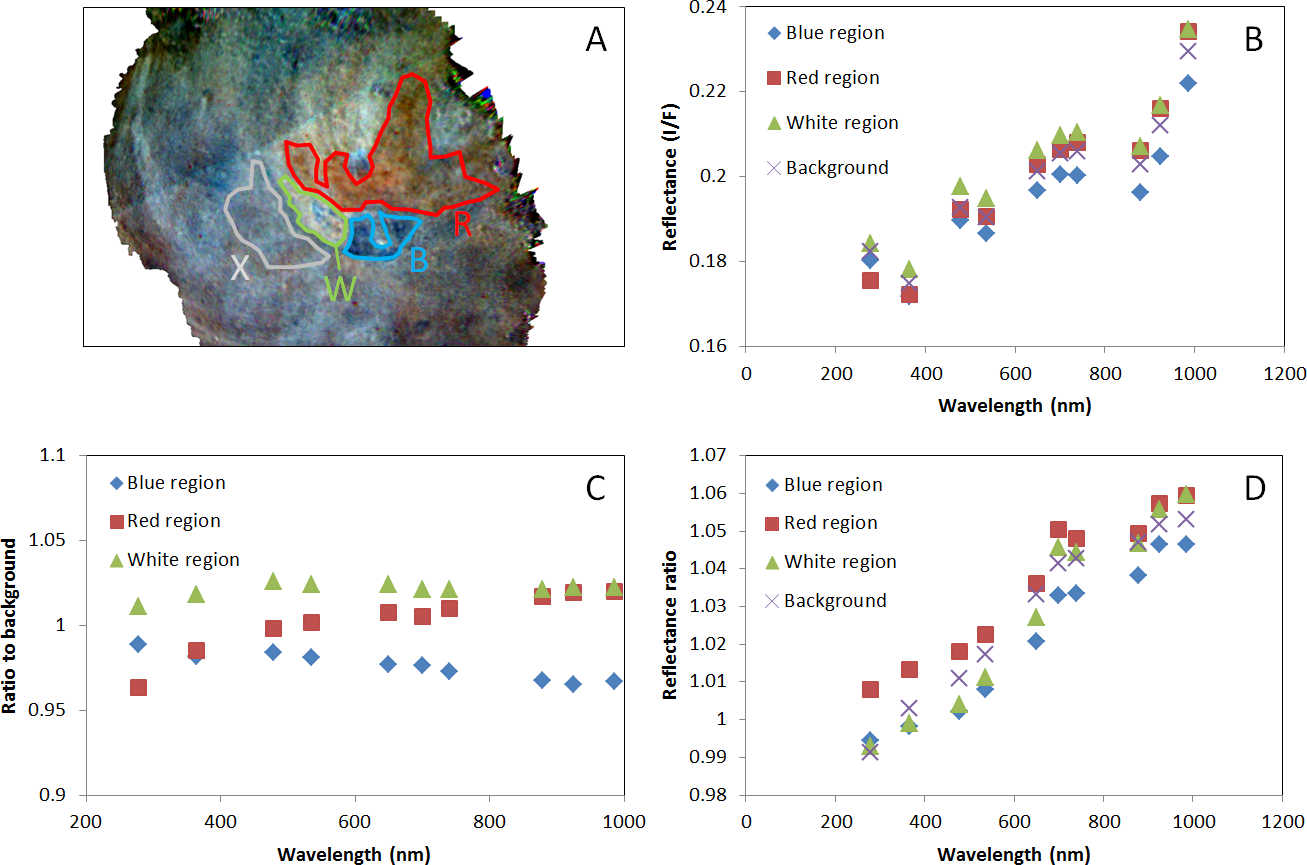}
\caption{Spectral properties of four regions in the Baetica region. ({\bf A}) Definition of the {\it Red} (R), {\it Blue} (B), {\it White} (W), and background (X) units in the color composite of set~1. ({\bf B}) Photometrically corrected reflectance in the low phase angle images. ({\bf C}) Reflectance relative to the background region in the low phase angle images. ({\bf D}) Phase reddening, defined as the ratio of the reflectances at low and high phase angle.}
\label{fig:spectra}
\end{figure}

In Fig.~\ref{fig:spectra} we further examine the spectral properties of the Baetica region. We define three units inside the crater complex (Fig.~\ref{fig:spectra}A) and compare their average spectra with that of a background unit outside the complex. These units are: Red for the crater interior that is orange in the photometrically corrected composites, Blue for the larger of the two dark blue terrains, and White for the white patch adjacent to the other units. We show the results for the images in set~1 because they offer a higher signal-to-noise (S/N) than those in set~2. When we plot the absolute reflectance as a function of wavelength (Fig.~\ref{fig:spectra}B) we find that the spectrum of all units is relatively featureless and weakly increases with wavelength, consistent with \citet{Mg12}. In this representation, the units are difficult to distinguish, both from each other and the background. However, differences are readily apparent when we plot the reflectance relative to that of the background (Fig.~\ref{fig:spectra}C). The advantage of using the relative reflectance is that it is insensitive to calibration errors and stray light (see also Sec.~\ref{sec:pca}). The Red unit is redder than the background over the full wavelength range, most strongly in the UV. The Blue unit is darker than the background over the full wavelength range and its relative reflectance spectrum has a blue slope. The White unit is brighter than the background over the full wavelength range. Its relative spectrum features a red slope below 400~nm, similar to that of the Red unit, but is flat at higher wavelengths; it appears to be a linear combination of the Red and Blue unit spectra. Its reflectance is the highest at all wavelengths. The fact that the reflectivity of a silicate particulate surface is known to increase with decreasing particle size \citep{AF67} suggests that the average particle size in the White unit is smaller than that in the other units. An analysis of set~2 confirms the results for set~1. Given the average phase angle difference between the two sets ($15.3^\circ$) we can search for evidence of spectral changes as a function of phase angle. When we divide the reflectance of each unit in set~2 (high phase angle) by that in set~1 (low phase angle), we find that all units exhibit phase reddening, i.e.\ an increase of the spectral slope with increasing phase angle (Fig.~\ref{fig:spectra}D). While the differences are very small, phase reddening appears to be strongest for the White unit and weakest for the Blue unit.

\subsection{Principal component analysis}
\label{sec:pca}

\begin{figure}
\centering
\includegraphics[width=7.5cm]{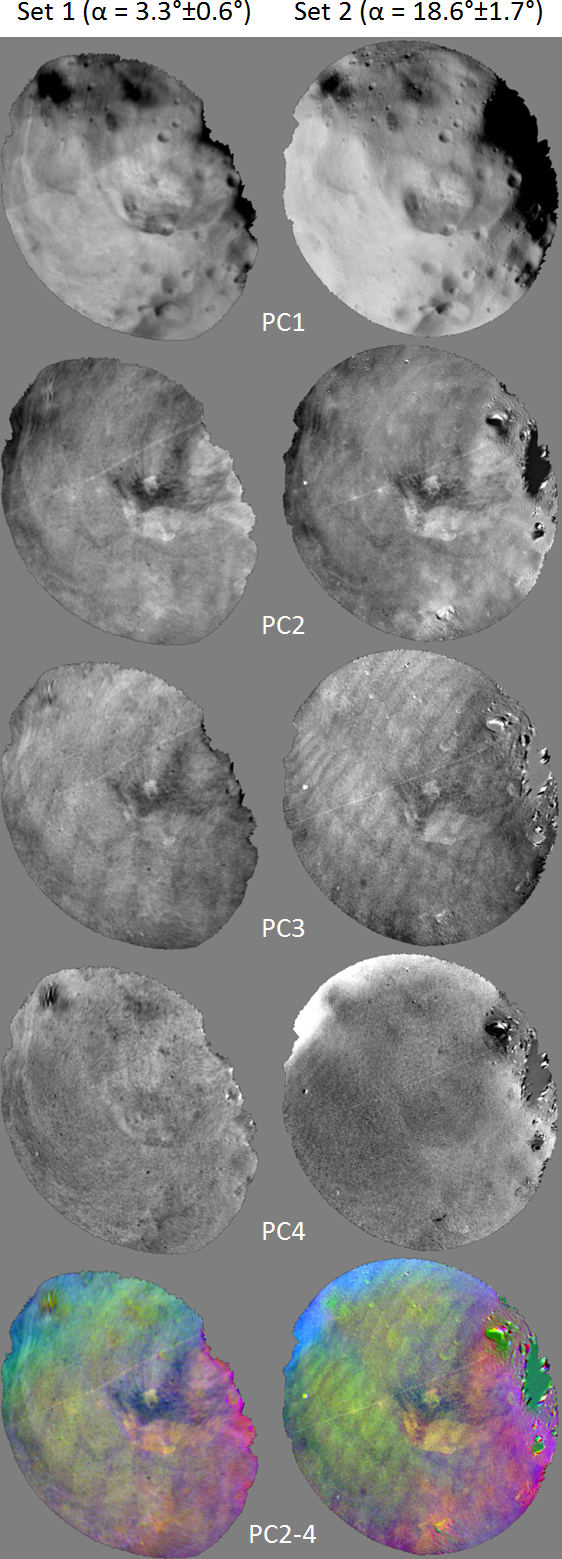}
\caption{Principal component analysis of two sets of orthorectified images, with low ({\bf left}) and high ({\bf right}) average phase angle ($\alpha$). Shown are the first four principal components (PC1-4). PC1 shows what correlates most over all bands, i.e.\ the brightness (including shadowing), whereas higher components show color variations over the surface. Also shown is a color composite of the higher components PC2 (red), PC3 (green), and PC4 (blue). The highest components contain only noise, and are not shown.}
\label{fig:pca}
\end{figure}

\begin{figure}
\centering
\includegraphics[width=\textwidth]{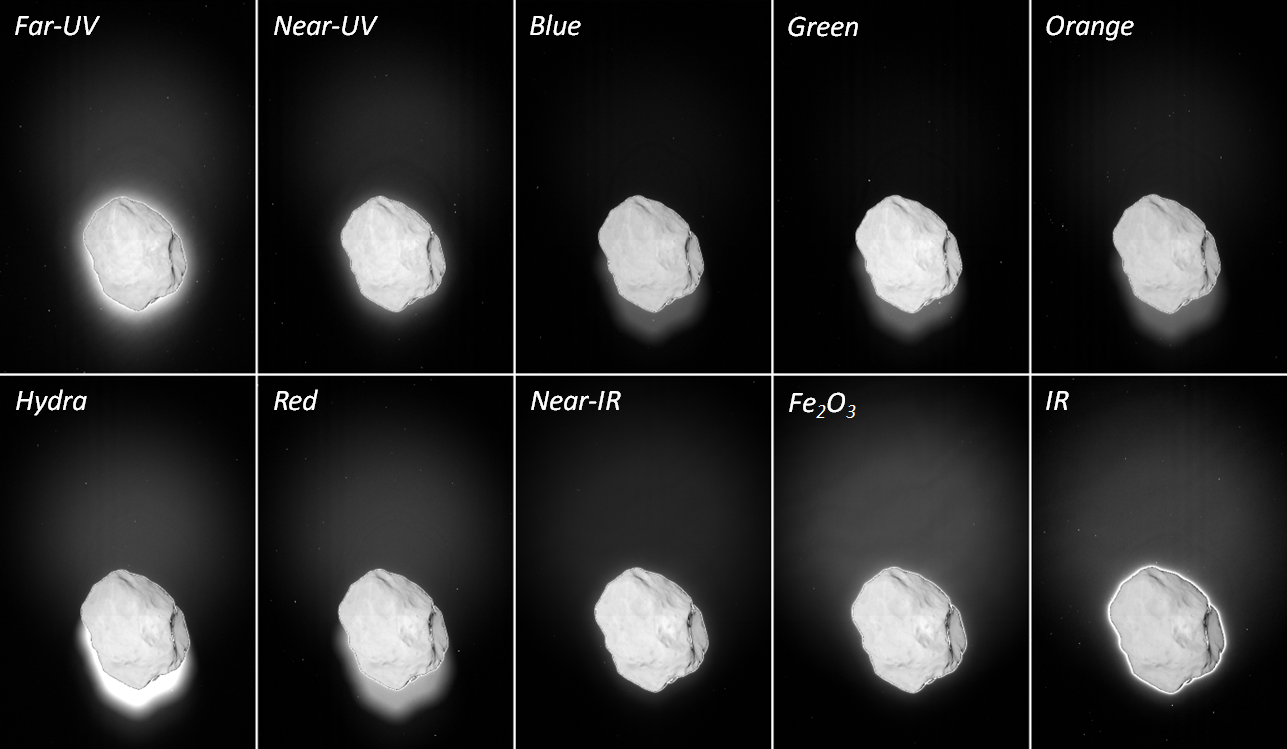}
\caption{Stray light and ghosts in the original images of set~1. The area around the asteroid itself is enhanced such that white is 5\% of the median asteroid signal and black is zero.}
\label{fig:stray_light}
\end{figure}

An alternative method to analyze the color images is the principal component analysis (PCA), with which one can separate brightness variations from color variations. The former are isolated in the first principal component (PC1), whereas the latter are isolated in the higher principal components. We consider color variations real if they are present in both data sets. PC2 has the effects of brightness removed and contains the dominant color variation. Higher components contain more subtle color variations in addition to calibration artifacts, image registration errors, and noise. Figure~\ref{fig:pca} shows the results of a PCA on our two sets of orthorectified images. PC1 shows what correlates most over all bands, i.e.\ the apparent brightness, both due to variations in albedo and shading. PC1 of set~2 is completely dominated by the effects of shadows and shading and explains 99.7\% of the total variance in the data. Shadows are weaker in set~1 because of the low phase angle ($3.3^\circ$) and we can clearly distinguish albedo variations. PC1 of set~1 explains 94.4\% of the variance. PC2 shows that the dominant color variations are small, explaining only 2.8\% and 0.14\% of the variance in set~1 and 2, respectively. PC3 shows surface features but also a diagonally banded pattern that is associated with periodic noise in the original images, and explains 1.4\% (set~1) and 0.045\% (set~2) of the variance. PC4 of set~1 shows very weak surface features that cannot be recognized in set~2, and explains only 0.46\% (set~1) and 0.027\% (set~2) of the variance. All color variations intrinsic to the surface are present in PC2-4; the higher components contain only noise and are not shown here. We can display all color variations in one image by constructing color composites of PC2-4 (Fig.~\ref{fig:pca}, bottom row). Color variations intrinsic to the surface are found in the center, in the Baetica crater complex. Large scale color gradients in the outer parts of the composites, like the blue upper left corner and the pink right side, are most likely due to stray light and image ghosts. This can be understood with help of Fig.~\ref{fig:stray_light}, which shows the stray light for each image in set~1. We distinguish several forms of stray light. The first one manifests itself as a fuzzy halo around the image of Lutetia, and is especially strong for the {\it Far-UV} and {\it IR} filters. The second is represented by a ghost image of the asteroid, offset by about 50~pixels from the asteroid image itself, in downward direction in Fig.~\ref{fig:stray_light}, and is strongest for the visible and near-IR filters, especially {\it Hydra} (7\%). Note that the ghost image adds signal to the asteroid image, and the reconstructed reflectance in Fig.~\ref{fig:spectra}B is too high by several percent for most filters, especially in the visible. Another form of stray light is a large, fuzzy spherical blob, with its center about 100~pixels above the asteroid in Fig.~\ref{fig:stray_light}. Its strength depends on the filter, being strongest for {\it Fe$_2$O$_3$} and virtually absent from {\it Blue} and {\it Green}. So the large scale color variations are probably artifacts. But to establish this beyond doubt requires accurate modeling of the stray light contribution, which is beyond the scope of this paper.

\begin{figure}
\centering
\includegraphics[width=8cm]{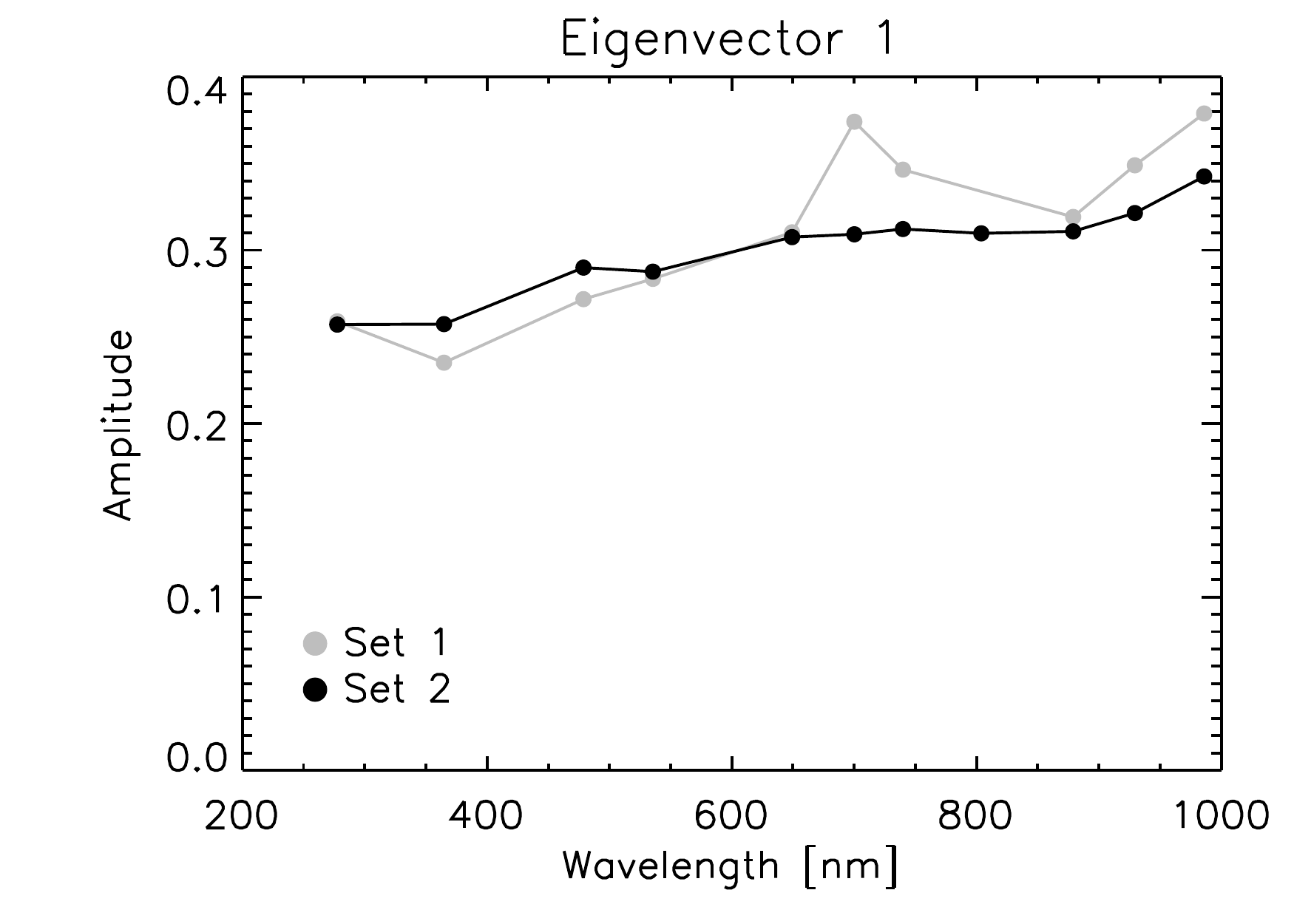}
\includegraphics[width=8cm]{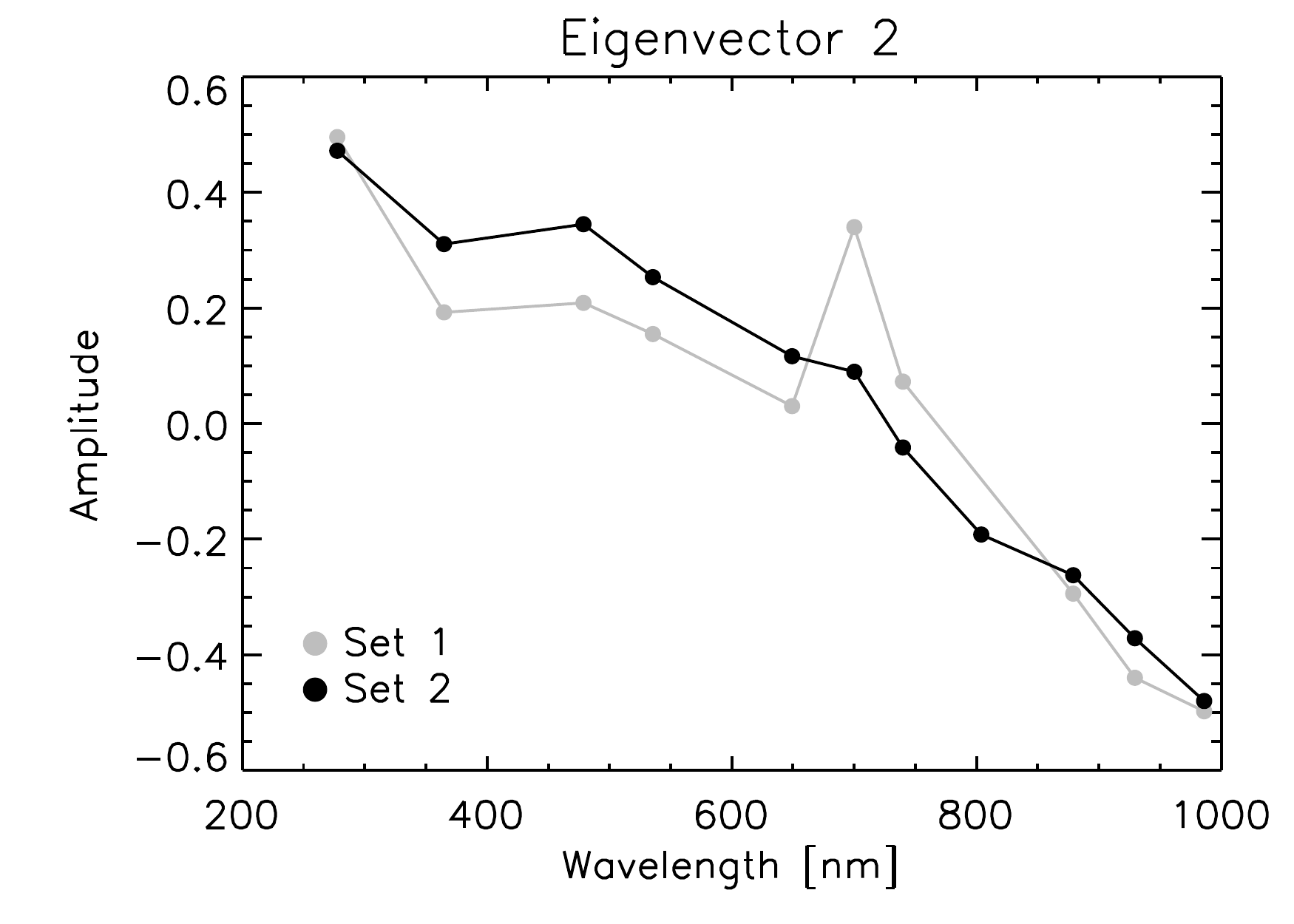}
\includegraphics[width=8cm]{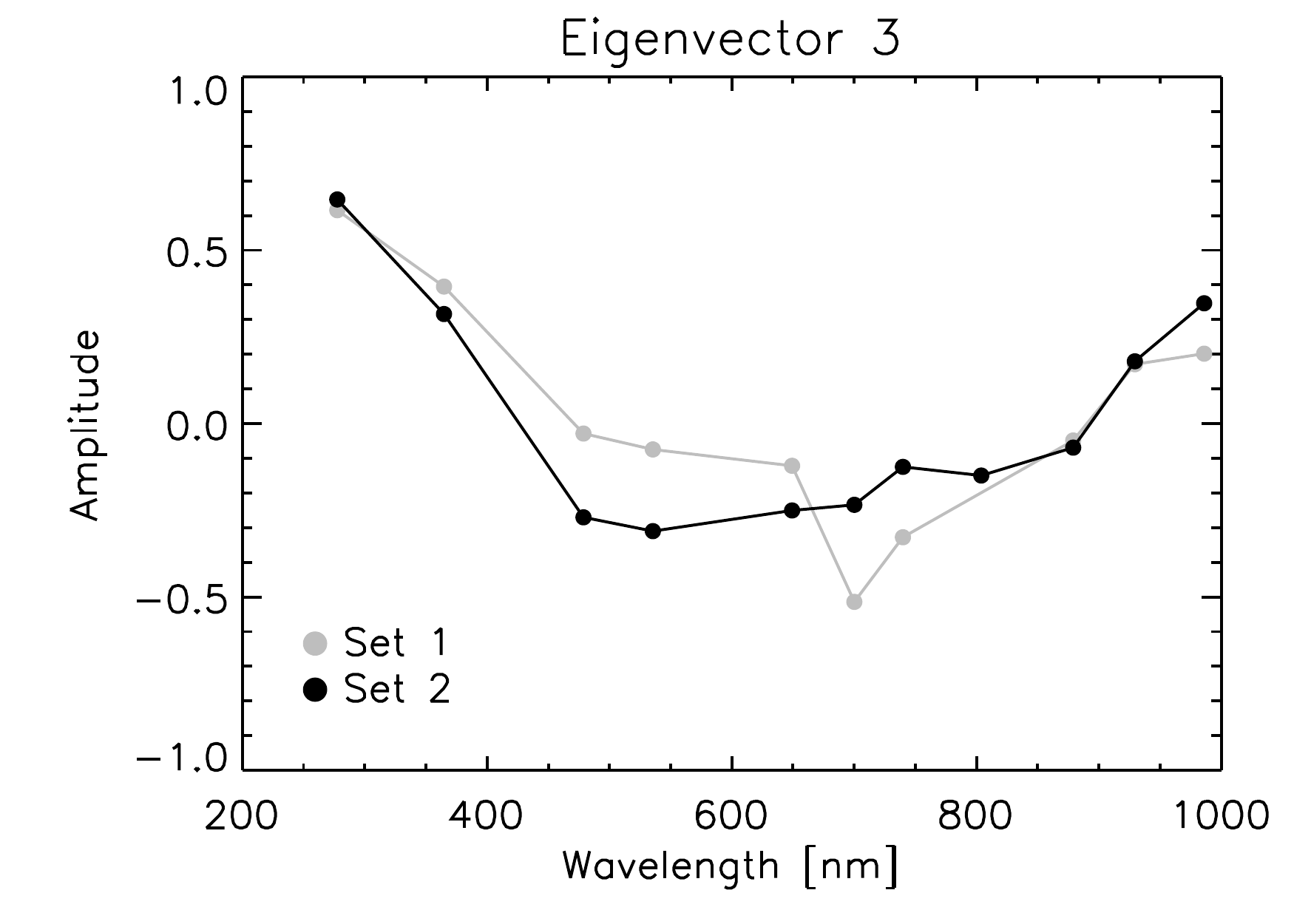}
\includegraphics[width=8cm]{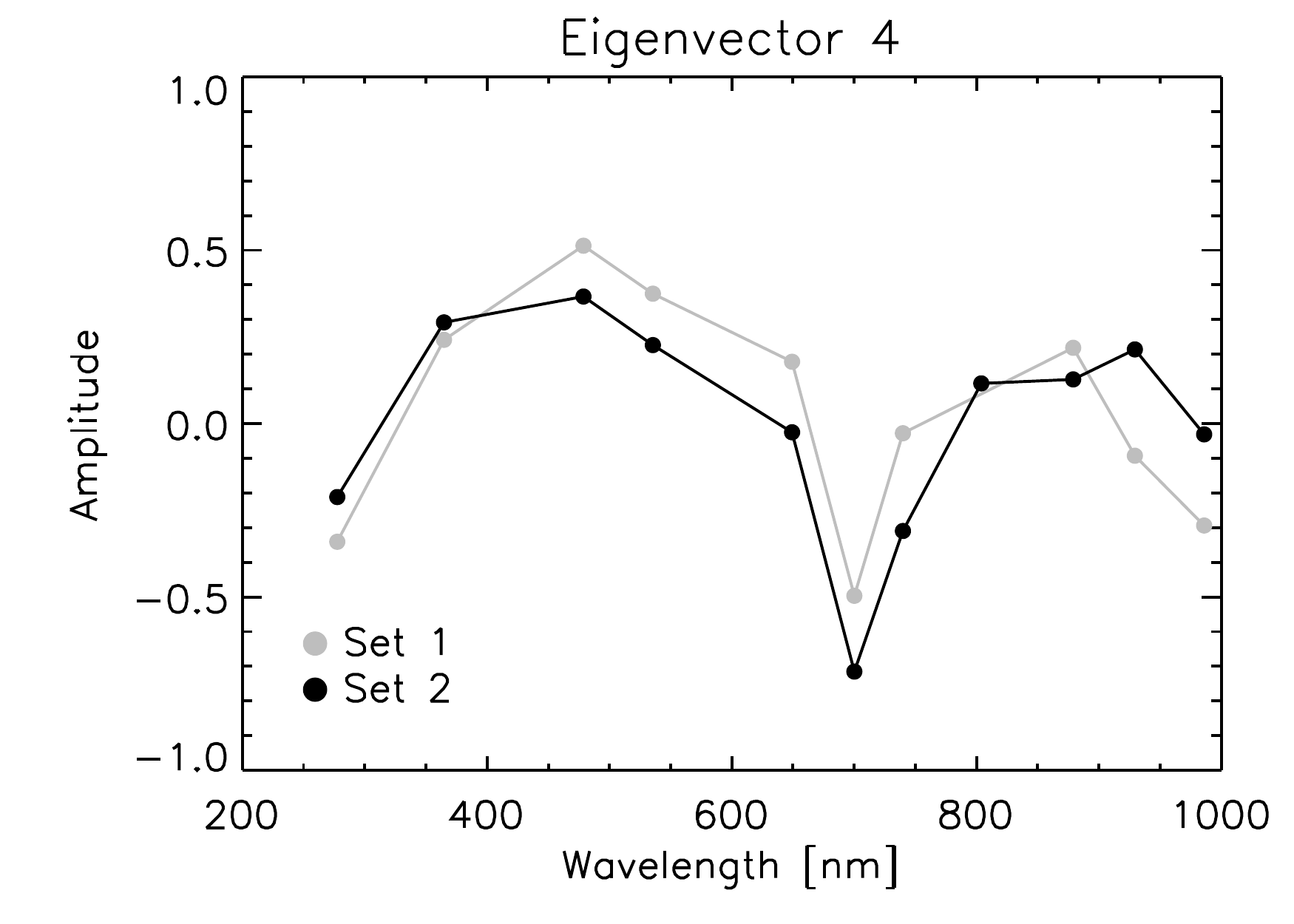}
\caption{The eigenvectors associated with the principal components in Fig.~\ref{fig:pca}.}
\label{fig:eigenvectors}
\end{figure}

While the principal component images show us where the color variations are located, the eigenvectors (Fig.~\ref{fig:eigenvectors}) show us the nature of the variations. There is one eigenvector for each PC. Again, we only consider features real if they are present in the eigenvectors of both sets, where we note that the eigenvectors of set~1 have a higher S/N. The first eigenvector is associated with PC1 and shows the average reflectance spectrum of the surface. It does not exhibit any significant absorption features and is very similar to the spectra in Fig.~\ref{fig:spectra}B. The second eigenvector shows the dominant color variation, i.e.\ the dominant type of deviation from the average spectrum. It has a featureless blue slope. This means that areas that are bright (dark) in the PC2 images in Fig.~\ref{fig:pca} are bluer (redder) than average. In the central Baetica region we find areas that are bright and dark, meaning bluer and redder. The third eigenvector shows a more complex shape: a blue slope from 250 to 450~nm, and a red slope from 900 to 1000~nm. As surface features are readily recognized in PC3, these spectral variations must be real, although we are unable to interpret them. Except for the periodic noise, PC3 is rather similar to PC2, and eigenvector~3 represents a small modulation of the spectrum associated with eigenvector~2. In particular, it signifies the stronger red slope in the UV for the Red unit in Fig.~\ref{fig:spectra}. The fourth eigenvector shows a distinct feature at 700~nm, which is probably real because it is present in the eigenvectors of both sets. However, the PC4 of both sets are too different to assume that it represents an actual absorption feature in the spectrum of Lutetia.

\subsection{Baetica region}
\label{sec:baetica}

\begin{figure}
\centering
\includegraphics[width=\textwidth]{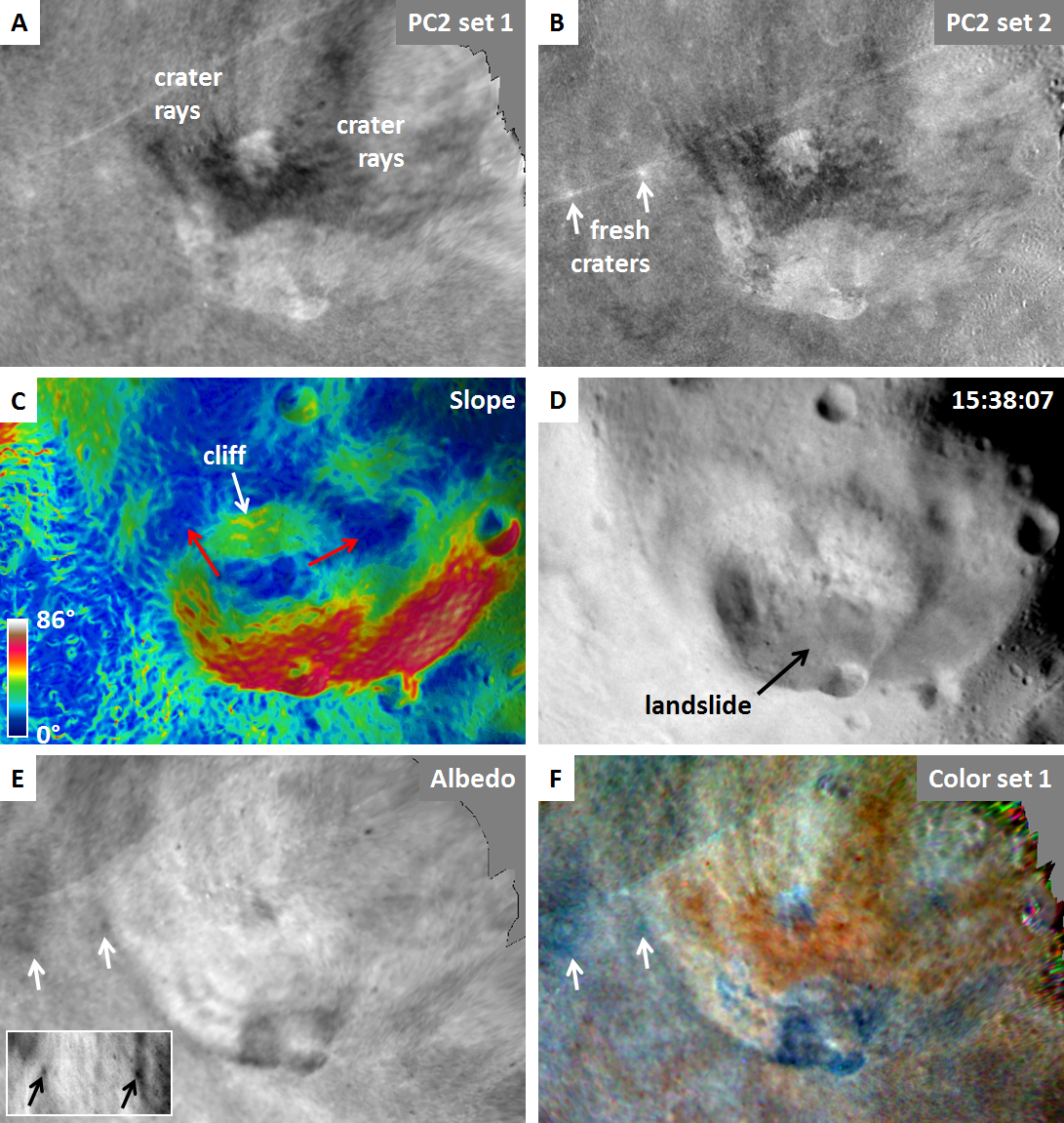}
\caption{Six views of the Baetica crater complex with surface features labeled. ({\bf A}) PC2 of set~1. ({\bf B}) PC2 of set~2. ({\bf C}) Physical slope, calculated with respect to a 47~km sphere. The red arrows indicate directions with a small gradient. ({\bf D}) {\it Green} image 15:38:07. ({\bf E}) albedo (opposition) image. ({\bf F}) False color image for set~1. The bright diagonal line in the PC2 images is an artifact associated with a gain difference for two parts of the CCD. The white arrows in (B), (E), and (F) point at two fresh craters that appear dark in {\it Orange} image 15:42:41 (E, inset).}
\label{fig:Baetica}
\end{figure}

The strongest color variations are present in the crater cluster in the Baetica region. Fortunately, as this area was in the center of the OSIRIS images the registration errors are minimal here. Figure~\ref{fig:Baetica} takes a closer look at this region. PC2 is ideally suited to study the color variations here as it is relatively free of noise and suffers less from artifacts in the shape model than the photometrically corrected images. This allows features to be detected with the highest possible spatial resolution, especially in set~2. Areas that are bright (dark) in PC2 are bluer (redder) than average. We verify the reality of features by comparing the PC2 of both sets (Fig.~\ref{fig:Baetica}A and B). The opposition image in Fig.~\ref{fig:Baetica}E is ideally suited to study variations in visible albedo. The crater cluster harbors terrains that are both bluer and redder. The bottom part of the crater cluster is the site of a major landslide, which is bluer and darker than average. The top part of the cluster is redder and brighter than average with streaks of material that are oriented radially to the cluster center. We identify these streaks as crater rays. Two sets of rays are separated by a small bright area that we identify as a promontory in a map of the physical slope (Fig.~\ref{fig:Baetica}C), identified as a rock outcrop by \citet{M12}. The distribution of rays roughly correlates with that of boulders, which also appear to avoid the promontory \citep{K12}. Below it there is another landslide ({\it Gallicum Labes}), which is also relatively blue. Left of the crater complex are two small spots that are bright in the PC2 of both sets. They are barely resolved in the OSIRIS image sets, but appear dark in an high-resolution {\it Orange} filter image taken near closest approach (Fig.~\ref{fig:Baetica}E, inset). They can also be discerned in the low phase angle false color composite as faint blue smudges (Fig.~\ref{fig:Baetica}F). These spots are probably small impact craters. The distinct color of small craters on S-type asteroids and the Moon tends to fade quickly due to space weathering, either due to induced color changes \citep{H94,S96} or mixing \citep{Pi12}. As some of these processes may also occur on E-type asteroids, we infer that the spots are fresh impact craters. The fact that the landslides are dark and blue like these small craters suggests that they are equally young. The landslides postdate the crater itself, which is estimated to be younger than 300 million years based on boulder counts \citep{K12}.

To explain the color distribution in the Baetica crater complex we envision the following scenario. An impact created the central crater and covered the crater floor with a relatively red material. We do not know whether this material is endogenous or exogenous. The surge that deposited this material was predominantly lateral and led to the formation of crater rays identified in PC2. It followed the local topography in a direction where the slope is minimal (indicated by red arrows in Fig.~\ref{fig:Baetica}C), avoiding obstacles like the promontory near the top of the complex, and did not extend beyond the steep crater wall at the bottom of the complex. After the impact the crater was modified by landslides, both on the steep crater wall and the promontory on the opposite side. They covered the reddish deposits with fresh, relatively dark and blue material from below the surface. Not easy to explain is the bright material on one side of the crater wall that has spectral characteristics intermediate to that of the red and blue deposits (the White unit in Fig.~\ref{fig:spectra}). Possibly, small landslides created a thin veneer of fine-grained fresh material on top of the mature deposit. Another explanation is that fine-grained fresh and mature materials were mixed by the combined action of impact and landslides. Given our conclusion that fresh material is relatively blue, one may expect mature material to be relatively red. However, the Red unit cannot be mature, being located inside the young Baetica cluster. It appears that the reddening experienced by this unit is not a consequence of maturity in terms of space weathering, but may derive instead from processes such as impactor material deposition or shock metamorphosis. In the PCA analysis, PC2 does not simply represent the degree of soil maturation. In PC2, bright equals fresh but dark does not necessarily equal mature.

\section{Discussion}
\label{sec:discussion}

Accurately orthorectifying the Lutetia images has proved to be an essential prerequisite for the study of the subtle color variations on the surface. Due to the stray light characteristics of the OSIRIS camera our study is constrained to color variations on small scales near the image center. Such variegation is easiest to detect at low phase angle, when shadows and shading are weak. These restrictions make the crater cluster in the Baetica region and its immediate surroundings an ideal study target. We confirm the Baetica region as a site of variegation \citep{Mg12}. A landslide in the crater complex and two small craters in the vicinity of the complex are both darker and bluer than the surrounding regolith. Given that these features were most likely created in recent events, we suggest that their blue color is due to a lack of space weathering. If indeed space weathering on Lutetia leads to reddening and brightening of the regolith, then this would be a surprising result. First identified on the moon, space weathering leads to reddening and darkening of the lunar regolith in the visible wavelength range \citep{P93}. The mechanism behind these changes is thought to be coating of regolith particles with vapor-deposited nanophase iron (see overview by \citealt{C04}). Color variations on the surface of S-type asteroids are commonly found and generally attributed to lunar-style space weathering. On S-type asteroids Gaspra and Ida crater ridges are brighter and bluer than average \citep{H94,S96}. Oddly, Eros has large albedo variations but only very small color variations \citep{B02}. Space weathering on Eros does not appear to lead to significant reddening for reasons not well understood \citep{C04}. Small Itokawa again follows the trend for S-type asteroids, with brighter areas being bluer and darker areas being redder \citep{S06}.

Space weathering manifests itself differently on other asteroid types. The surfaces of C-type asteroids have not been studied at very high resolution due to a lack of dedicated spacecraft missions. \citet{C99} attributed a lack of color and normal reflectance variations on C-type asteroid Mathilde to an absence of space weathering alteration of the optical properties of the surface. Subtle color variations were identified on E-type asteroid {\v S}teins by \citet{S10}, though not confirmed by \citet{Le10}. The interior of the large crater on the south pole appears relatively blue, and the authors suggested a link with space weathering. While V-type asteroid Vesta is surprisingly colorful, it does not display obvious signs of space weathering. The dominant color variation involves the depth of the 1~\textmu m absorption band, and is associated with the large impact basin on the south pole \citep{R12}. Minor color variations involve the spectral slope and small absorption features in the visible wavelength range. \citet{Pi12} suggested that Vesta space weathering does not alter the optical properties of the regolith but merely thoroughly mixes it.

Lutetia is an unusual type of asteroid and we do not know how space weathering would manifests itself on the surface. \citet{V09} performed irradiation experiments on an enstatite chondrite to simulate space weathering. The authors argued that Lutetia is a plausible parent body for this meteorite class. They found spectral changes (reddening and darkening) that were only minor compared to those observed for ordinary chondrites at similar levels of irradiation. Space weathering is the most natural explanation for some of the color variations that we uncovered. The slightly bluer color of fresh, unweathered material is consistent with the irradiation experiments. However, this material is also darker than the more mature surroundings, which is contrary to the experiments and expectations for lunar-style space weathering. Possibly, the albedo of fresh deposits is lower because of a larger grain size. Alternatively, large impacts may redden the regolith through an unknown process (shock metamorphism?) and landslides and small impacts may expose a blue sub-surface unit. Vesta-like space weathering could then mix these two materials into a regolith with average optical properties. \citet{C11} cited Lutetia's flat spectrum and its spectral homogeneity as observed by VIRTIS as evidence for a lack of color changes induced by space weathering. Although its spatial resolution is much lower than that of OSIRIS (the two small craters in Fig.~\ref{fig:Baetica} were not resolved), VIRTIS was able to study color variations over the entire visible surface. The scarcity of color variations that it found suggests that weathering-induced color changes on the surface of Lutetia occur faster than for S-type asteroids.

We confirm the small degree of phase reddening found by \citet{Mg12}. Phase reddening is commonly observed for surfaces of terrestrial planets but defies easy explanation. One form of reddening results from the increase in optical path length with increasing phase angle for a surface of semi-transparent particles for which the absorption coefficient decreases with wavelength \citep{AF67}. However, the roughness of the surface and that of the particles themselves also play a role; monotonous reddening can be expected for a surface that is rough on the scale of the particles, i.e.\ a typical planetary regolith \citep{S14}. While in this sense Lutetia meets the expectations, the phase angle separation of our two image sets ($\sim 15^\circ$) is too small to permit any inferences beyond that a regolith may be present.

\section*{Acknowledgements}

The authors thank an anonymous referee for useful comments.

%\end{linenumbers}

\bibliography{Lutetia}

\newpage
\clearpage

\begin{table}
\centering
\caption{Overview of the Lutetia narrow-angle camera (NAC) images acquired on 10 July 2010 that are analyzed in this paper. Image sets~1 ($n=10$) and 2 ($n=11$) are referred to as low- and high-phase angle sets, respectively. The time is that reported in the image file name. The true start of the acquisition time (UTC) is 19 sec later. The distance and phase angle are listed as reported in the image header. We excluded an {\it Ortho} image from set~1 because of a camera shutter error.}
\vspace{5mm}
\begin{tabular}{cllccc}
\hline
Set & Time       & Filter        & Distance & Phase angle & Resolution \\
    & (hh:mm:ss) &               & (km)     & (degree)    & (m/pixel) \\
\hline
  & 15:26:36 & {\it Orange}      & 16,484 & 0.15 & 306 \\
1 & 15:30:13 & {\it Far-UV}      & 13,317 & 2.7 & 247 \\
1 & 15:30:23 & {\it Near-UV}     & 13,161 & 2.8 & 244 \\
1 & 15:30:32 & {\it Blue}        & 13,038 & 2.9 & 242 \\
1 & 15:30:40 & {\it Green}       & 12,919 & 3.0 & 239 \\
1 & 15:30:49 & {\it Orange}      & 12,793 & 3.2 & 237 \\
1 & 15:30:57 & {\it Hydra}       & 12,667 & 3.4 & 235 \\
1 & 15:31:06 & {\it Red}         & 12,545 & 3.5 & 232 \\
1 & 15:31:29 & {\it Near-IR}     & 12,208 & 3.9 & 226 \\
1 & 15:31:37 & {\it Fe$_2$O$_3$} & 12,092 & 4.1 & 224 \\
1 & 15:31:45 & {\it IR}          & 11,978 & 4.2 & 222 \\
2 & 15:37:41 & {\it Far-UV}      & 6,977 & 15.9 & 129 \\
2 & 15:37:50 & {\it Near-UV}     & 6,860 & 16.4 & 127 \\
2 & 15:37:58 & {\it Blue}        & 6,753 & 16.9 & 125 \\
2 & 15:38:07 & {\it Green}       & 6,634 & 17.4 & 123 \\
2 & 15:38:15 & {\it Orange}      & 6,529 & 17.9 & 121 \\
2 & 15:38:23 & {\it Hydra}       & 6,418 & 18.5 & 119 \\
2 & 15:38:32 & {\it Red}         & 6,308 & 19.0 & 116 \\
2 & 15:38:40 & {\it Ortho}       & 6,197 & 19.6 & 114 \\
2 & 15:38:48 & {\it Near-IR}     & 6,097 & 20.2 & 113 \\
2 & 15:38:56 & {\it Fe$_2$O$_3$} & 5,993 & 20.8 & 111 \\
2 & 15:39:04 & {\it IR}          & 5,893 & 21.4 & 109 \\
\hline
\end{tabular}
\label{tab:data_sets}
\end{table}

%\newpage
%\clearpage

\begin{table}
\centering
\caption{Characteristics of the OSIRIS NAC filters used during the Lutetia campaign \citep{K07}. The number in the first column appears in the image file name.}
\vspace{5mm}
\begin{tabular}{clcc}
\hline
\# & Name              & Wavelength & Bandwidth \\
     &                   & (nm)       & (nm) \\
\hline
15   & {\it Far-UV}      & 269        & 54 \\
16   & {\it Near-UV}     & 360        & 51 \\
84   & {\it Blue}        & 481        & 75 \\
83   & {\it Green}       & 536        & 62 \\
82   & {\it Orange}      & 649        & 85 \\
87   & {\it Hydra}       & 701        & 22 \\
88   & {\it Red}         & 744        & 64 \\
51   & {\it Ortho}       & 805        & 41 \\
41   & {\it Near-IR}     & 882        & 66 \\
61   & {\it Fe$_2$O$_3$} & 932        & 35 \\
71   & {\it IR}          & 989        & 38 \\
\hline
\end{tabular}
\label{tab:filters}
\end{table}

\newpage
\clearpage

\newpage
\clearpage

\newpage
\clearpage

\newpage
\clearpage

\newpage
\clearpage

\newpage
\clearpage
\thispagestyle{empty}

\newpage
\clearpage

\newpage
\clearpage

\newpage
\clearpage

\end{document}